\documentclass[traditabstract]{aa}
\usepackage{graphicx}
\usepackage{natbib}
\usepackage{txfonts}
\begin{document}
\title{Structure function of the UV variability of Q0957+561}


\author{L. J. Goicoechea\inst{1}, V. N. Shalyapin\inst{2}, 
        R. Gil-Merino\inst{3} \and A. Ull\'an\inst{4}}

\institute{Departamento de F\'{\i}sica Moderna, Universidad de Cantabria, 
	       Avda. de Los Castros s/n, 39005 Santander, Spain\\
             \email{goicol@unican.es}
\and
		 Institute for Radiophysics and Electronics, National Academy of Sciences of Ukraine, 
             12 Proskura St., Kharkov 61085, Ukraine\\
             \email{vshal@ire.kharkov.ua}
\and
             Instituto de F\'{\i}sica de Cantabria (CSIC-UC), Avda. de Los Castros s/n, 39005 
             Santander, Spain\\
             \email{gilmerino@ifca.unican.es}
\and
             Robotic Telescopes Group, Centro de Astrobiolog\'{\i}a (CSIC-INTA), associated to the 
             NASA Astrobiology Institute, Ctra de Ajalvir, km 4, 28850 Torrej\'on de Ardoz, Madrid, 
             Spain\\
             \email{ullanna@inta.es}}

\titlerunning{Structure function of the UV variability of Q0957+561}
\authorrunning{Goicoechea et al.}


\abstract{We present a detailed structure function analysis of the UV variability of 
\object{Q0957+561}. From optical observations in the 2005$-$2007 seasons, we constructed normalized 
structure functions of the quasar luminosity at restframe wavelengths $\lambda \sim$ 2100 \AA\ 
and $\lambda \sim$ 2600 \AA. Old optical records (1995$-$1996 seasons) also allow the structure 
function to be obtained at $\lambda \sim$ 2100 \AA, but 10 years ago in the observer's frame. These 
three structure functions are then compared to predictions of both simple and relatively 
sophisticated (incorporating two independent variable components) Poissonian models. We do not 
find clear evidence of a chromatic mechanism of variability. From the recent data, $\sim$ 100-d 
time-symmetric and $\sim$ 170-d time-asymmetric flares are produced at both restframe wavelengths. 
Taking into account measurements of time delays and the existence of an EUV/radio jet, 
reverberation is probably the main mechanism of variability. Thus, two types of EUV/X-ray 
fluctuations would be generated within or close to the jet and later reprocessed by the disc gas 
in the two emission rings, at $\lambda \sim$ 2100 \AA\ and $\lambda \sim$ 2600 \AA. The $\sim$ 100-d 
time-symmetric shots are also responsible for most of the $\sim$ 2100 \AA\ variability detected in 
the old experiment (10 years ago). However, there is no evidence of asymmetric shots in the old 
UV variability. If reverberation is the involved mechanism of variability, this could mean an
intermittent production of high-energy asymmetric fluctuations. The old records are also consistent
with the presence of very short-lifetime ($\sim$ 10 d) symmetric flares, which may represent  
additional evidence of time evolution. Despite these exciting findings, we cannot rule out the
possibility of unfortunate gaps in the old light curves (or a relatively short monitoring period)
and very short-timescale systematic noise. We also discuss the quasar structure that emerges from 
the variability scenario.}


\maketitle

\section{Introduction}

Structure functions for optical/UV variability of quasars allow investigation of processes 
that cause intrinsic variations and the properties of intrinsic flares 
\citep[e.g.][]{Kaw98,Cid00,Van04,deV05,Wil08}, the composition of intervening systems
\citep[e.g.][]{Lew96,Sch03}, and the nature of the intergalactic medium \citep[e.g.][]{Haw02}.
The most recent and comprehensive study of ensemble structure functions of non-lensed quasars
has been presented by \citet{Wil08}. In this work, about 2500 quasars at a median redshift of 
$\sim$ 2 were classified into six different groups according to their black hole mass and 
continuum luminosity. For each group, \citet{Wil08} derived the square root of the noise-less 
structure function for the $ugriz$ bands. They detected the well-known anticorrelation between 
luminosity and variability, as well as a correlation between variability and black hole mass
\citep[see also][]{Wol07}. Taking these findings into account, most of the variations at 
restframe time lags $\leq$ 1 year are probably intrinsic to the quasars. These might be due 
to differences in accretion rate. The analysis by \citet{LiC08} also supports this last claim.
However, \citet{Wol07} and \citet{Wil08} point out the absence of a clear correlation between 
variability and black hole mass for restframe time lags $\leq$ 100 d. Therefore, short
timescale variations could not be related to changes in accretion rate. 

The Wilhite et al. dataset did not allow for a measurement of the variability of very 
bright and massive quasars. Moreover, the mechanism of variability might differ from local to 
distant quasars, and other physical properties (apart from luminosity and black hole mass) 
may play a role (e.g., circumnuclear activity, presence of a jet, X-ray activity, etc.). Thus, 
detailed structure functions of well-characterised individual quasars are important tools for 
understanding different quasar populations. With respect to the individual sources, for example, 
\citet{Cid00} analysed the members of a sample of optically selected nearby quasars \citep[$z 
<$ 0.4; see][]{Giv99}. They used Poissonian models, and derived flare lifetimes and other 
parameters of interest. A Poissonian model describes physical scenarios in which the luminosity 
is due to the superposition of a variable component and a constant background (the nonflaring 
part of the emissivity). The variable component is made by the superposition of flares, having 
a given profile and occurring at random times \citep{Cid00}. \citet{Cid00} reported the poor 
sensitivity of their fits to phenomenological models (flare profiles), with the exception of 
exponentially decaying flares. These asymmetric flares led to poor fits. \citet{Col01a} also 
studied optical/UV structure functions of 13 local AGN (Seyfert 1 galaxies) with very good 
time coverage and resolution. The flare lifetimes (using symmetric triangular flares and 
certain lag intervals) were $\tau \sim$ 5$-$100 days, with the higher mass AGN having larger
variability timescales.

Many previous studies focused on the initial logarithmic slope of the structure function. For
example, using the square root of the structure function, a measured initial slope can be 
compared to predictions of possible physical mechanisms of quasar variability 
\citep[e.g.][]{Kaw98,Haw02}. The initial logarithmic slope $\beta$ should be less than or 
equal to 0.5 for square and exponentially decaying flares (Poissonian phenomenological models), 
whereas symmetric triangular flares produce a steep slope $\beta >$ 0.5 \citep{Cid00,Goi08}. 
One can also discuss simple models of the reverberation scenario, i.e., accretion disc flares 
induced by variable EUV/X-ray irradiation. The cellular-automaton model is able to reproduce a 
slope $\beta \sim$ 0.4$-$0.5 \citep{Min94,Kaw98}, but cellular-automaton simulations neglect 
hydrodynamical effects. These might be basic ingredients in the reverberation scenario. 
The one-dimensional hydrodynamical simulations by \citet{Man96} led to time-symmetric flares,
i.e., flares having symmetric rise and decay. These time-symmetric flares can account for 
steep growths with slopes above 0.5. The starburst model (supernova explosions) produces an 
initial slope $\beta \geq$ 0.7 \citep{Are97,Kaw98}, and the microlensing model (extrinsic 
variability caused by collapsed objects passing close to the sight line toward a quasar) leads 
to a shallow slope $\beta \sim$ 0.2$-$0.3 \citep{Haw02}. For non-lensed quasars, the slope of 
the square-root noise-less structure function over restframe lags $\geq$ 100 d is roughly 0.5 
\citep[e.g.][]{Cid00,Wil08}. On the other hand, local AGN show a variety of initial slopes 
over restframe lags $\leq$ 100 d: $\beta \sim$ 0.3$-$0.8 \citep{Col01a}. For timescales 
$\leq$ 60 d, \citet{Col01a} also found that optical and UV slopes are consistent with each 
other. This supports a common variability mechanism. 

Over the last 10$-$15 years, several gravitationally lensed quasars have been monitored more 
or less regularly. Some of them offer a unique opportunity to study the origin of the 
intrinsic signal of quasars, since their intrinsic fluctuations are repeated in two or more 
lensed components with certain time delays and magnitude offsets \citep[e.g.][]{Sch92}. 
Optical frames ($g$ and $r$ bands) of the double quasar \object{Q0957+561} were taken with 
the Apache Point Observatory (APO) 3.5 m telescope during the 1995 and 1996 seasons. These 
produced accurate light curves that basically include intrinsic variations \citep{Kun97}. 
\citet{Kaw98} studied the square root of the structure function of \object{Q0957+561} using 
the APO $g$-band light curves (photometric magnitudes), finding a shallow logarithmic slope 
of $\sim$ 0.35 over restframe lags $\leq$ 200 d. Unfortunately, these authors did not 
subtract the observational noise from the photometric measurements 
\citep[e.g.,][]{Sim85,Cid00,Col01a}, which has a significant effect on the measured 
variations at the shortest lags. \citet{Gil01} found an initial slope of the square-root 
noise-less structure function of \object{Q0957+561} in the $R$ band $\beta >$ 0.5 (timescale 
$<$ 100 d). Very recently, \citet{Foh08} separated intrinsic from extrinsic variations in 
\object{Q1004+4112}, and measured a slope of $\sim$ 0.5 over restframe lags $\leq$ 1 year 
(intrinsic structure function in the $r$ band). From $r$-band data of \object{Q0909+532}, 
\citet{Goi08} also estimated a steep initial slope $\beta \sim$ 1 over restframe lags $\leq$
100 d. No simple model was able to accurately reproduce the shape of the structure function 
of the intrinsic luminosity of \object{Q0909+532}. 

The first lensed quasar \object{Q0957+561} \citep{Wal79} is probably the best-studied lens 
system. This has been investigated in several spectral regions, including radio, IR, optical, 
UV and X-ray wavelengths. The radio maps of both components showed the presence of radio 
cores and $\sim 0\farcs1$ jets \citep{Por81,Gor84,Gor88,Gar94,Haa01}. \citet{Hut03} reported 
evidence for EUV ($\sim$ 1100 \AA) activity along the radio jets. Apart from EUV emission 
associated with the jet, \citet{Hut03} also found EUV emission within a radius of $0\farcs3$, 
which is associated with a circumnuclear environment. Taking into account the redshift of 
the quasar ($z$  = 1.41), optical observations in the $g$ and $r$ bands correspond to middle 
ultraviolet (MUV) emission ($\sim$ 2100$-$2600 \AA). Hence, fluctuations in the observed
optical magnitude trace variations in the UV luminosity of the source. \object{Q0957+561} is
a very bright and massive object, with both $\lambda L_{\lambda}$(1350 \AA) and 
$\lambda L_{\lambda}$(3000 \AA) exceeding 10$^{46}$ erg s$^{-1}$, and a black hole mass of
2$-$3 $\times$ 10$^9$ M$_\odot$ \citep{Pen06}. The double quasar \object{Q0957+561} is also 
a X-ray bright source \citep[e.g.][]{Cha00}, but a possible X-ray jet has not yet been 
resolved. 

Liverpool Quasar Lens Monitoring (LQLM) is a long-term project to follow the variability of 
gravitationally lensed quasars with the Liverpool robotic telescope \citep{Goi08}. The first 
phase of this project (LQLM I) included a monitoring programme of \object{Q0957+561} in the 
$gr$ optical bands. The new LQLM I light curves of \object{Q0957+561} (2005$-$2007 seasons) 
indicated the absence of significant extrinsic variability \citep{Sha08}. In addition, using 
the time delays (between quasar components and between optical bands) from the APO and LQLM I 
records, \citet{Sha08} suggested that most prominent variations in \object{Q0957+561} are 
very probably due to reverberation in the accretion disc around the supermassive black hole 
\citep[see also][]{Col01b}. In Section 2, the new LQLM I dataset is used to accurately trace 
the shapes of the square-root noise-less structure functions of the quasar UV luminosity. 
These shapes are closely linked to the nature of the UV fluctuations (see here above), so we 
investigate the mechanism(s) of variability at two different restframe wavelengths: $\lambda 
\sim$ 2100 \AA\ ($g$ band) and $\lambda \sim$ 2600 \AA\ ($r$ band). By comparing the new shape 
with the old (APO dataset) at $\lambda \sim$ 2100 \AA, we also check for time evolution of the 
mechanism(s) of variability. In Section 3, we present our conclusions and discuss the quasar 
structure (temperature profile and source sizes) that emerges from the favoured variability 
scenario. 

\section{Structure function analysis}

\begin{figure}
\centering
\includegraphics[angle=0,width=9cm]{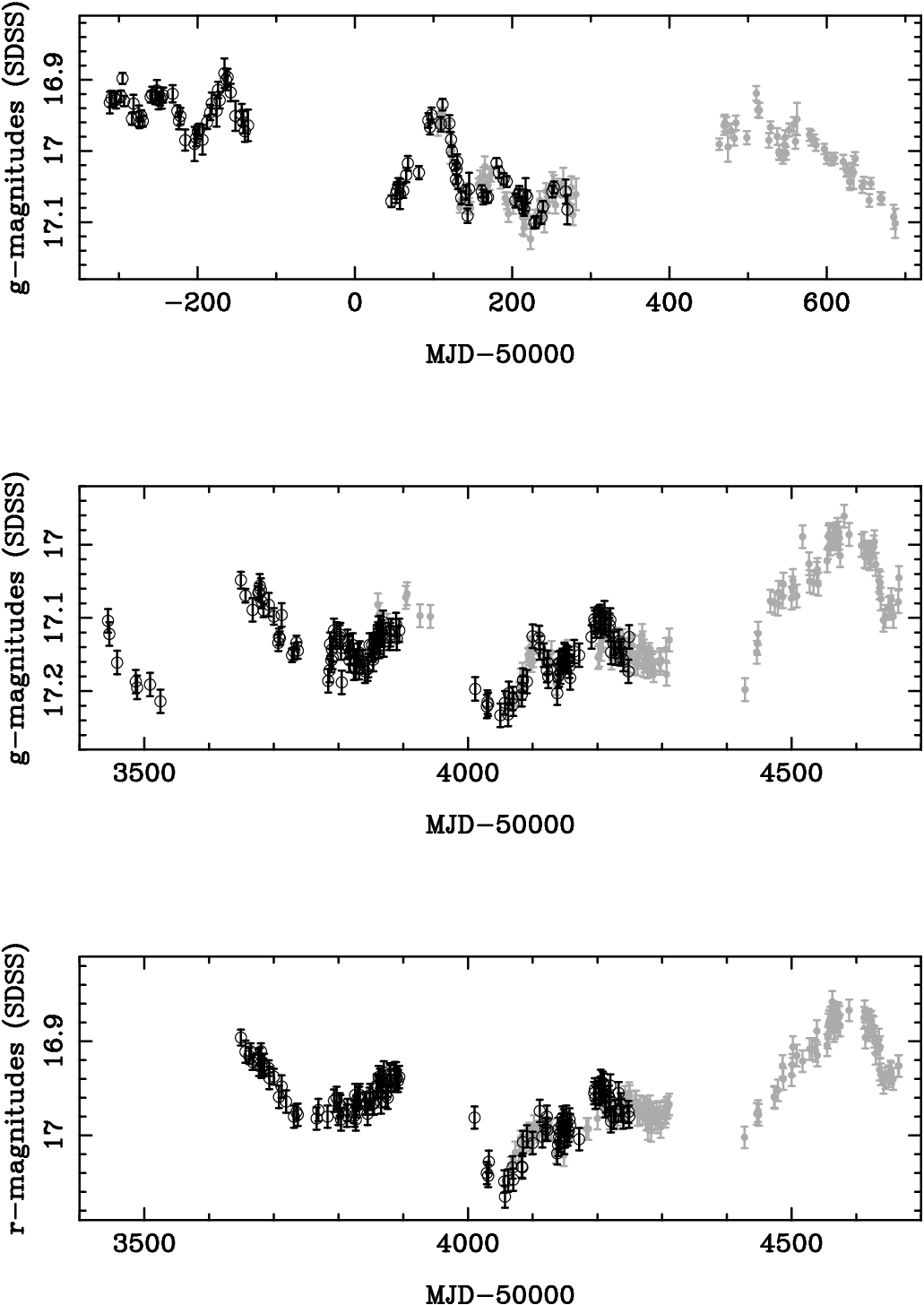}
\caption{APO and LQLM I combined light curves. While the light curves of Q0957+561B (open 
circles) are unchanged, the Q0957+561A records (filled circles) are shifted by the 417-d 
time delay and optimal magnitude offsets (producing optimal overlaps). Top: APO $g$-band. 
Middle: LQLM I $g$-band. Bottom: LQLM I $r$-band. We note that all magnitudes are expressed
in the SDSS photometric system.}
\label{combi}
\end{figure}

The APO and LQLM I light curves of the double quasar \object{Q0957+561} do not show evidence 
of extrinsic variability \citep{Kun97,Sha08}. Thus, we directly obtain the noise-less 
structure function of the intrinsic luminosity at a given restframe wavelength \citep[instead 
of one structure function for each component, a combined record and the corresponding 
structure function are made, e.g.,][]{Kaw98}. The combined records that we use in this paper
are depicted in Fig.~\ref{combi}. There is no standard form of the structure function, but 
different approaches to the problem. For example, while magnitudes, $SF(m)$, are often used in 
optical astronomy \citep[e.g.][and references therein]{Wil08}, monochromatic fluxes or 
luminosities, $SF(F)$ or $SF(L)$, are more relevant in radio astronomy or to accurately 
compare with models of variability \citep[e.g.][]{Sim85,Cid00,Col01a}. The structure function 
$SF(L)$ at restframe lag $\Delta \tau$ is estimated through the averaged sum
\begin{equation}
SF(L) = (1/2N) \sum_{i,j} [(10^{- 0.4 m_j} - 10^{- 0.4 m_i})^2 - \overline{\sigma}_i^2 - 
\overline{\sigma}_j^2] ,
\end{equation}
where $m$ are magnitudes, $\overline{\sigma} = 0.921 \times 10^{- 0.4 m} \sigma$, $\sigma$ are 
photometric uncertainties, and the sum includes $N$ pairs verifying $\tau_j - \tau_i \sim 
\Delta \tau$. Here, $L = 10^{- 0.4 m}$ are monochromatic luminosities in convenient units and 
$\tau$ are restframe times \citep[e.g.,][]{Cid00,Goi08}. This $SF(L)$ describes typical 
luminosity variabilities at different restframe lags. We normalize the original structure 
function to the luminosity variance and then take the square-root for convenience, i.e., we
analyse the normalized structure function $f = [SF(L)]^{1/2} / \sigma(L)$.    

\begin{figure}
\centering
\includegraphics[angle=-90,width=9cm]{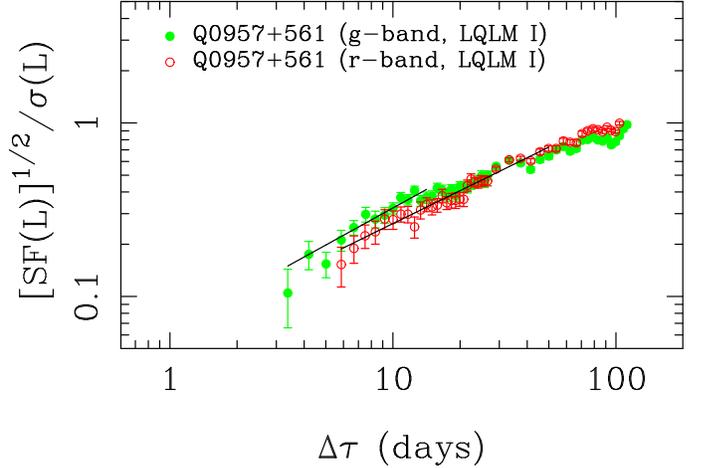}
\caption{Normalized structure functions of Q0957+561 from Liverpool telescope (LQLM I) records.
These structure functions of the UV luminosities at $\sim$ 2100 \AA\ ($g$ band; filled circles) 
and $\sim$ 2600 \AA\ ($r$ band; open circles) are accurately described from very short 
restframe lags ($\sim$ 1$-$10 days) to lags when $[SF(L)]^{1/2} / \sigma(L) \sim$ 1 (asymptotic 
behaviours). The initial growths are consistent with slopes $\beta \sim$ 0.6$-$0.8 (solid lines).}
\label{sfgr}
\end{figure}

Restframe lags substantially below the restframe duration of the records are considered in 
the analysis. In the initial selection, we take $\Delta \tau \leq P/4(1 + z)$, where $P$ 
is the duration of each whole combined record (see Fig.~\ref{combi}). Later, only lags before 
reaching the asymptotic behaviours ($f \leq$ 1) are taken into account. In Fig.~\ref{sfgr}, 
using LQLM I observations, the normalized structure function at $\lambda \sim$ 2100 \AA\ ($g$ 
band; filled circles) is compared to $f$ at $\lambda \sim$ 2600 \AA\ ($r$ band; open circles). 
Both growths seem to be consistent with each other. With respect to the initial logarithmic 
slopes, in Fig.~\ref{sfgr} we also show two fits $f = A (\Delta \tau)^{\beta}$ with 
$\hat{\chi}^2 = \chi^2/N_{dof}$ values close to 1 ($N_{dof}$ is the number of degrees of 
freedom). The fits over time intervals 3$-$14 d ($g$ band) and 6$-$50 d ($r$ band) lead to 
$\beta$ = 0.71 $\pm$ 0.08 and $\beta$ = 0.63 $\pm$ 0.02 (1$\sigma$ intervals), respectively. 
These initial slopes disagree with the prediction of the cellular-automaton disc-instability 
model, but roughly agree with the time-symmetric flares that appear in the one-dimensional 
hydrodynamical simulations (see Introduction). In principle, the starburst model could also 
account for the measured slope at $\sim$ 2100 \AA\ ($g$ band). The microlensing slope $\beta 
\sim$ 0.2$-$0.3 \citep{Haw02} is strongly inconsistent with the LQLM I data of 
\object{Q0957+561}, which is not at all surprising. We are studying intrinsic fluctuations, so 
microlensing does not play any role.

\begin{figure}
\centering
\includegraphics[angle=-90,width=9cm]{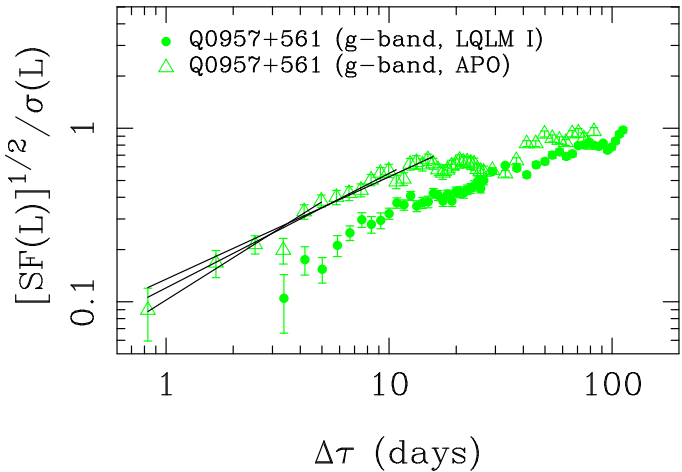}
\caption{Normalized structure functions of Q0957+561 from Apache Point Observatory (APO) and 
Liverpool telescope (LQLM I) records. The two structure functions of the luminosity at $\sim$ 
2100 \AA\ (APO = open triangles, LQLM I = filled circles) correspond to two experiments separated
by $\sim$ 10 years, i.e., $\sim$ 4 years in the quasar restframe. The initial logarithmic slopes 
(see solid lines) of the APO trend vary from $\beta \sim$ 0.8 over lags $\leq$ 5 d to $\beta \sim$ 
0.6 over lags $\leq$ 16 d.}
\label{sfgg}
\end{figure}

In Fig.~\ref{sfgg}, the APO (open triangles) and LQLM I (filled circles) structure functions of 
\object{Q0957+561} at $\sim$ 2100 \AA\ have significantly different initial growths. This may be a 
consequence of evolution in the variability scenario, since both experiments (APO and LQLM I) are 
separated by $\sim$ 1500 d in the quasar restframe. In fact, the oldest (APO) brightness records 
seem to incorporate relatively short fluctuations (which would not be present in the LQLM I light 
curves) that would generate the differences between initial growths. The solid lines in 
Fig.~\ref{sfgg} fit the APO behaviours in three time intervals ($\hat{\chi}^2 \sim$ 1). Their 
slopes are $\beta$ = 0.81 $\pm$ 0.12 at $\Delta \tau \leq$ 5 d, $\beta$ = 0.66 $\pm$ 0.06 at 
$\Delta \tau \leq$ 11 d, and $\beta$ = 0.59 $\pm$ 0.04 at $\Delta \tau \leq$ 16 d (1$\sigma$ 
intervals). 

\begin{table*}
\begin{minipage}[t]{\columnwidth}
\caption{Solutions for the main Poissonian models.}
\label{fits}
\centering
\renewcommand{\footnoterule}{}  
\begin{tabular}{llcccc}
\hline\hline
Model\footnote{SQF = square flares, EDF = exponentially decaying flares, STF = symmetric triangular 
flares, and SBF = starburst flares.} & Observed structure function\footnote{Normalized structure 
functions (see main text) from Liverpool telescope (LQLM I) and Apache Point Observatory (APO) data.} 
& $\hat{\chi}^2$ & $w_1$\footnote{All measurements are 1$\sigma$ intervals.} 
& $\tau_1$$^c$ (d) 
& $\tau_2$$^c$ (d)\\                  
\hline
SQF(1)+STF(2) &$f_{LQLM\ I}$(2100\ \AA) &1.21 &0.266 $^{+0.016}_{-0.020}$ &32.6 $^{+1.4}_{-2.4}$ &175.8 $^{+7.2}_{-6.6}$\\ 
              &$f_{LQLM\ I}$(2600\ \AA) &0.65 &0.170 $^{+0.028}_{-0.016}$ &33.2 $^{+4.2}_{-1.6}$ &122.7 $^{+4.2}_{-3.9}$\\   
              &$f_{APO}$(2100\ \AA) &1.44 &0.266 $^{+0.024}_{-0.024}$ &11.0 $^{+1.4}_{-1.2}$ &107.1 $^{+6.6}_{-6.0}$\\ 
SQF(1)+SBF(2) &$f_{LQLM\ I}$(2100\ \AA) &1.19 &0.706 $^{+0.092}_{-0.664}$ &151.4 $^{+10.8}_{-142.2}$ &135 $^{+161}_{-29}$\\ 
              &$f_{LQLM\ I}$(2600\ \AA) &0.99 &0.630 $^{+0.080}_{-0.082}$ &88.2 $^{+4.0}_{-6.0}$ &353 $^{+108}_{-52}$\\   
              &$f_{APO}$(2100\ \AA) &1.78 &0.758 $^{+0.042}_{-0.046}$ &99.4 $^{+8.6}_{-6.8}$ &32 $^{+6}_{-6}$\\ 
EDF(1)+STF(2) &$f_{LQLM\ I}$(2100\ \AA) &1.25 &0.902 $^{+0.042}_{-0.268}$ &180.4 $^{+14.0}_{-62.8}$ &90.6 $^{+107.7}_{-26.1}$\\ 
              &$f_{LQLM\ I}$(2600\ \AA) &0.73 &0.418 $^{+0.048}_{-0.054}$ &154.8 $^{+27.6}_{-30.4}$ &97.2 $^{+9.6}_{-7.2}$\\   
              &$f_{APO}$(2100\ \AA) &1.80 &0.856 $^{+0.032}_{-0.030}$ &135.6 $^{+14.0}_{-12.4}$ &10.8 $^{+1.5}_{-1.5}$\\
STF(1)+STF(2) &$f_{LQLM\ I}$(2100\ \AA) &1.55 &0.204 $^{+0.020}_{-0.018}$ &24.3 $^{+2.4}_{-2.7}$ &162.3 $^{+6.3}_{-5.7}$\\ 
              &$f_{LQLM\ I}$(2600\ \AA) &0.80 &0.182 $^{+0.046}_{-0.036}$ &36.0 $^{+5.4}_{-5.1}$ &123.9 $^{+5.7}_{-5.1}$\\   
              &$f_{APO}$(2100\ \AA) &1.31 &0.270 $^{+0.018}_{-0.020}$ &10.8 $^{+0.9}_{-0.9}$ &108.0 $^{+6.3}_{-5.7}$\\  
STF(1)+SBF(2) &$f_{LQLM\ I}$(2100\ \AA) &1.18 &0.052 $^{+0.586}_{-0.010}$ &11.7 $^{+177.3}_{-2.1}$ &287 $^{+13}_{-192}$\\ 
              &$f_{LQLM\ I}$(2600\ \AA) &0.72 &0.608 $^{+0.110}_{-0.098}$ &112.5 $^{+6.0}_{-8.4}$ &167 $^{+35}_{-40}$\\   
              &$f_{APO}$(2100\ \AA) &1.64 &0.646 $^{+0.036}_{-0.036}$ &114.3 $^{+8.4}_{-7.8}$ &33 $^{+5}_{-4}$\\  
\hline
\end{tabular}
\end{minipage}
\end{table*}

In order to discuss the mechanism(s) for (M)UV variability, and to quantify the spectral and time 
evolution of that mechanism(s), we compare the observed shapes of the structure functions 
(Figs.~\ref{sfgr}$-$\ref{sfgg}) with predictions of Poissonian models. In a first level of 
complexity (simplest models), we use three phenomenological models: square flares (SQF), 
exponentially decaying flares (EDF), and symmetric triangular flares (STF), as well as starburst 
flares (SBF) produced by supernova explosions \citep{Are97,Cid00}. For example, EDF is a simple 
model to describe asymmetric flares, i.e.,  rapid rises and slow declines, or slow rises and rapid 
declines (the structure function cannot distinguish between both variants). STF is a rough 
description of the hydrodynamical simulations by \citet{Man96} and a relatively good approach to 
the X-ray shots found by \citet{Neg94}. Each model is characterised by a shape function $s(\tau)$, 
$f = [s(\tau)]^{1/2}$, and these shape functions appear in Appendix B of \citet{Cid00} and Eq. 
(17) of \citet{Are97}. The 4 level-1 models of $f$ only include one free parameter: flare lifetime 
$\tau$ (e.g., $\tau = T$ for a square flare of duration $T$ and $\tau = 5 t_{sg}$ for a supernova 
explosion, where $t_{sg}$ is the time when the supernova remnant reaches the maximum of its 
radiative phase). In the Poissonian framework, we can consider more complex schemes. Thus, in a 
second level of complexity, the luminosity is assumed to be due to the superposition of a constant 
background and two independent variable components. It can easily be shown that $f = [w_1 
s_1(\tau_1) + (1 - w_1) s_2(\tau_2)]^{1/2}$, where $w_1$ is the ratio between the variance of the 
first fluctuating component and the total variance, and $s_1$ and $s_2$ are the shape functions of 
the two independent components.

The possibility that AGN variability is caused by several independent processes was suggested in 
previous work \citep[e.g.][]{Kaw98,Col01a}. Here, using 10 level-2 models of $f$ (SQF+SQF, SQF+EDF,
SQF+STF, SQF+SBF, EDF+EDF, EDF+STF, EDF+SBF, STF+STF, STF+SBF, and SBF+SBF), we also explore this 
possibility. Each of these 10 possible combinations is characterised by three free parameters: 
$w_1$, and two lifetimes $\tau_1$ and $\tau_2$ (see above). To evaluate the quality of the fits
obtained with the full set of 14 models, we analyse the $\hat{\chi}^2$ (reduced chi-square) values. 
For an acceptable fit, $\chi^2$ is expected to be in the range $N_{dof} \pm 2 (2N_{dof})^{1/2}$ 
(allowing $\chi^2 - N_{dof}$ differences of up to two standard deviations of the $\chi^2$ 
distribution), which implies 0.6 $\leq \hat{\chi}^2 \leq$ 1.4 ($N_{dof} \sim$ 45). As expected 
from the measured slopes, the five best solutions in terms of $\hat{\chi}^2$ are associated with 
(level-2) models incorporating either STF, or SBF, or both of them. However, only one out of our 42 
fits to $f_{LQLM\ I}$(2100\ \AA), $f_{LQLM\ I}$(2600\ \AA), and $f_{APO}$(2100\ \AA) (3 observed 
shapes $\times$ 14 models) gives $\hat{\chi}^2 \leq$ 1.4. This result means that the uncertainties 
are slightly underestimated or alternatively, the models do not describe all details. The 
statistical uncertainties in each structure function, see Eq. (1), are computed as the standard 
deviations of the means (averaged sums) for the different time lag bins. Although we use a very 
popular estimator of uncertainties, \citet{Col01a} noted that not all pairs of data in a given bin 
are independent. To address this problem, they multiplied their errors by 2$^{1/2}$. We adopt the 
Collier \& Peterson's perspective and fit again the observed structure functions.

After slightly enlarging the error bars, we concentrate on level-2 models consisting of STF/SBF
and anything else. Among these 7 models of $f$, we select those giving $\hat{\chi}^2 <$ 2 for
the three observed structure functions (main Poissonian models). The set of solutions is presented 
in Table~\ref{fits}. With regard to the spectral behaviour, four out of the 5 models in 
Table~\ref{fits} are able to accurately and simultaneously describe $f_{LQLM\ I}$(2100\ \AA) and 
$f_{LQLM\ I}$(2600\ \AA), i.e., they fit both LQLM I shapes acceptably well (0.6 $\leq \hat{\chi}^2 
\leq$ 1.4). The solutions for the first model (SQF+STF) indicate that symmetric triangular flares 
at $\sim$ 2100 \AA\ have a lifetime longer than the duration of STF at $\sim$ 2600 \AA. However, 
this is puzzling with the accretion disc paradigm. Similar lifetimes are expected in a 
reverberation scenario. Moreover, $\tau$(2100 \AA) $<$ $\tau$(2600 \AA) in a disc local-instability 
scenario, e.g., the thermal timescale is proportional to $(R^3/GM)^{1/2}$, where $M$ is the mass of 
the central black hole and $R \propto \lambda^{4/3}$ is the emission radius \citep[e.g.,][and 
references therein]{Goi08}. The solutions for the second model (SQF+SBF) incorporate supernova 
explosions with $t_{sg}$(2100 \AA) $<$ $t_{sg}$(2600 \AA). However, the timescale $t_{sg}$ is 
exlusively related to the circumstellar density and the total energy released in each explosion 
\citep[e.g.,][]{Are97}, so a chromatic timescale is not the expected result. 

\begin{figure}
\centering
\includegraphics[angle=-90,width=9cm]{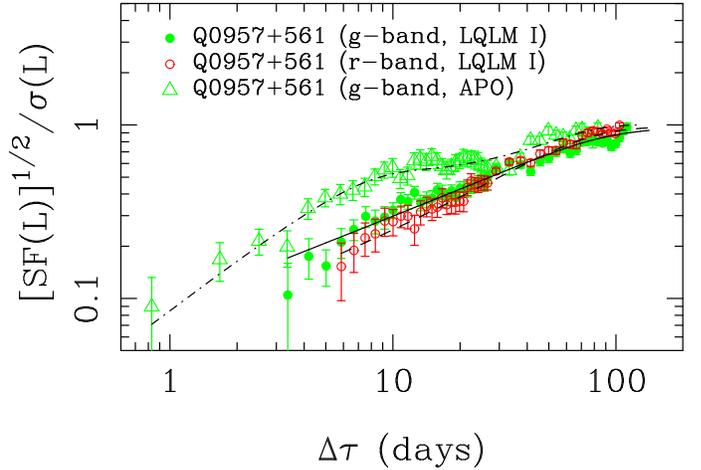}
\caption{Observed structure functions and adopted solutions. For the LQLM I experiment, we show 
EDF+STF laws with parameters: $w_1$ = 0.72, $\tau_1$ = 168 d, and $\tau_2$ = 105 d ($g$ band; solid
line), and $w_1$ = 0.40, $\tau_1$ = 168 d, and $\tau_2$ = 105 d ($r$ band; dashed line). SBF+STF 
laws with the above parameters are also able to reproduce the LQLM I shapes. For the APO data in 
the $g$ band, we display a STF+STF law with parameters: $w_1$ = 0.27, $\tau_1$ = 11 d, and $\tau_2$ 
= 105 d (dashed-dotted line).}
\label{sffits}
\end{figure}

The solutions for the EDF+STF and STF+SBF models are the most interesting ones (LQLM I data). There
is a clear degeneracy between EDF and SBF, which seem to work in a similar way. Hence, the 
solutions for the third and fifth models are interpreted as evidence in favour of the coexistence 
of $\sim$ 100-d time-symmetric flares and longer time-asymmetric shots (see Table~\ref{fits}). Both 
kinds of flares at both wavelengths can be mostly due to reverberation, i.e., two types of 
EUV/X-ray variation in the vicinity of the accretion disc axis that are reprocessed by two annuli 
of the disc gas. Measurements of time delays \citep{Kun97,Col01b,Sha08}, and the existence of an 
EUV jet \citep{Hut03} and a bright X-ray source \citep{Cha00}, support this physical scenario. From 
Table~\ref{fits}, $\tau_{asym}$ = 168 d and $\tau_{sym}$ = 105 d is a good compromise between the
results at the two wavelengths and for the two flare asymmetric profiles (exponentially decaying 
and starburst profile). Although the relative variances depend on wavelength, there is no need to 
invoke some mechanism other than reverberation. While the $\sim$ 2100 \AA\ variability is 
consistent with the asymmetric flares producing most of the variance at this wavelength ($w_{asym}$
= 72\%), the variance at $\sim$ 2600 \AA\ would be mainly due to the symmetric flares ($w_{asym}$ = 
40\%). This chromaticity in the relative variances is associated with the difference between time 
coverages, gaps and artifacts in the $g$ ($\sim$ 2100 \AA) and $r$ ($\sim$ 2600 \AA) bands (see 
middle and bottom panels in Fig.~\ref{combi}). The LQLM I $g$-band combined record lasts longer 
than the LQLM I combined curve in the $r$ band, so it contains a prominent decline that is not 
present in the shorter record. Besides the longer time coverage, the $g$-band curve fills a gap 
corresponding to the peak of an important event. The presence of a few artifacts could also play a 
role. In Fig.~\ref{sffits}, the LQLM I shapes are compared with the adopted solutions: $w_{asym}$ 
= 0.72, $\tau_{asym}$ = 168 d, and $\tau_{sym}$ = 105 d for $f_{LQLM\ I}$(2100\ \AA), and 
$w_{asym}$ = 0.40, $\tau_{asym}$ = 168 d, and $\tau_{sym}$ = 105 d for $f_{LQLM\ I}$(2600\ \AA). 

In spite of the enlargement of error bars and the use of relatively sophisticated modelling, no 
model leads to an acceptable global solution at both wavelengths and both epochs. The APO data are 
only consistent, in terms of $\hat{\chi}^2$, with the STF+STF model. The corresponding solution 
includes $\sim$ 100-d time-symmetric flares and other family of very short-lifetime ($\tau \sim$ 10 
d) symmetric flares. This last kind of fluctuations is responsible for about 1/4 of the total 
variance. The dominant $\sim$ 100-d shots are also detected in the LQLM I experiment, so the main 
source of these flares do not evolve over decades (in the observer's frame). Nevertheless, the 
absence of longer asymmetric flares (APO data) is a mystery. This could be due to unfortunate gaps 
in the light curves, time coverage of the monitoring, etc, or simply reflects the evolution of the 
central engine. Assuming that UV asymmetric flares are triggered by EUV/X-ray fluctuations 
(reverberation in the accretion disc; see above), we would be dealing with intermittency in the 
generation of high-energy asymmetric fluctuations (within or close to the jet). As we commented 
here above, the structure function $f_{APO}$(2100\ \AA) agrees with the presence of very 
short-timescale shots, which are not detected from the new LQLM I records. These fluctuations may 
be caused by some kind of observational (systematic) noise or alternatively, they might be 
related to an episode of very short-timescale activity inside the disc, the jet or other region. 
In Fig.~\ref{sffits}, the adopted solution (STF+STF model with parameters $w_1$ = 0.27, $\tau_1$ = 
11 d, and $\tau_2$ = 105 d) fits $f_{APO}$(2100\ \AA) reasonably well.

\section{Conclusions and discussion}

We present a novel and rigorous analysis of the structure function of the UV variability of the 
gravitationally lensed quasar \object{Q0957+561}. New Liverpool telescope data \citep[2005$-$2007 
seasons;][]{Sha08} allow us to construct normalized structure functions of the quasar luminosity at
two restframe wavelengths: $\lambda \sim$ 2100 \AA\ ($g$ band data) and $\lambda \sim$ 2600 \AA\ 
($r$ band data). Old Apache Point Observatory records in the $g$ band \citep[1995$-$1996 
seasons;][]{Kun97} are also used to check the possible time evolution of the variability at $\sim$ 
2100 \AA. The observed shapes of the structure functions are compared to predictions of a large set
of Poissonian models. This set of models includes the simplest and well-known ones, consisting of 
only one variable component \citep[][and references therein]{Cid00}, as well as hybrid models 
incorporating two independent variable components.

Several hybrid (or level-2) models are able to account for both Liverpool telescope structure 
functions (see Table~\ref{fits}). Some of them contain flares with unrealistic profile (square 
flares) and lead to solutions that are difficult to interpret. Fortunately, we also find reasonable
solutions in which $\sim$ 100-d time-symmetric and $\sim$ 170-d time-asymmetric flares are produced
at both restframe wavelengths. Exponentially decaying and starburst flares (and very probably 
other time-asymmetric shots) work in a similar way. Therefore, the good behaviour of the starburst 
ingredient does not necessarily implies the existence of supernova explosions, but the production 
of highly asymmetric shots. What about the mechanism of intrinsic variability?. The old and very 
recent $gr$ light curves of \object{Q0957+561} led to time delays between quasar components and 
between optical bands that mainly support a reverberation scenario \citep{Col01b,Sha08}. Thus, 
reverberation would be the main mechanism of variability. The presence of an EUV/radio jet 
\citep[e.g.,][]{Gar94,Hut03} and a bright X-ray source \citep{Cha00} also suggests the viability of 
this mechanism: two types of EUV/X-ray fluctuations that are generated within or close to the jet, 
and later reprocessed by two rings of the disc (each ring corresponds to a different restframe 
wavelength). On the other hand, one can also justify both kinds of flare profile. For example, the
cellular-automaton model produces asymmetric shots \citep[e.g.,][]{Kaw98}, and hydrodynamical 
simulations lead to symmetric flares \citep{Man96}. 

The $\sim$ 100-d time-symmetric shots seem to be also responsible for most of the $\sim$ 2100 \AA\
variability detected in the Apache Point Observatory experiment, but there is no evidence of 
asymmetric shots in the old UV variability. This absence of asymmetric flares may be due to gaps in
the light curves, a relatively short monitoring period, etc. Alternatively, it could means an
evolution of the central engine, i.e., intermittent production of high-energy asymmetric 
fluctuations. The Apache Point Observatory structure function is also consistent with the presence
of very short-lifetime ($\sim$ 10 d) symmetric flares. This kind of flare might be caused by 
observational systematic noise, or perhaps, represent additional evidence for time evolution. 
Our results do not support a previous claim for the possible starburst origin of some events in the
old $g$-band light curves \citep{Ull03}. Despite the presence of two twin events (one in each 
quasar component) with an anomalous delay \citep{Goi02}, the associated shot probably occurred at 
the base of the jet or in the circumnuclear region, but it was not originated by a supernova 
explosion.

Very recently, several studies have showed evidence that optical/UV variability of quasars on 
restframe timescales $>$ 100 d is mainly driven by variations in accretion rate 
\citep[e.g.,][]{Wol07,Are08,LiC08,Wil08}. Here we are discussing the UV variability of 
\object{Q0957+561} at restframe lags $\leq$ 100 d. \object{Q0957+561} is a very bright and massive
object, and this population could not be studied by \citet{Wil08}. The first lensed quasar has also
an EUV/radio jet (and important X-ray activity; see above), so high-energy variations in the 
surroundings of the disc axis and their reverberation are possible. For example, on timescales 
below 100 days, the optical variations of the local Seyfert galaxy \object{NGC 5548} are related to 
its X-ray variations \citep{Cze99}. Hence, part of the optical variability of this AGN (timescales 
$<$ 100 days) could be explained by X-ray reprocessing. Chromatic delays for the local Seyfert 
galaxy \object{NGC 7469} also reveal a reverberation scenario \citep{Col99}. \citet{Are08} reported 
on an illustrative example of mixed variability. The local quasar \object{MR 
2251-178} has been monitored simultaneously in X-rays and optical bands. All spectral regions were 
significantly variable, and the fluctuations were clearly correlated. \citet{Are08} indicated that 
pure reprocessing of X-rays cannot account for both $\sim$ 100-d and $\sim$ 500-d timescale optical 
variability. They claimed that two distinct mechanisms produce the variability: accretion rate 
variations plus reverberation, and the shortest timescale optical events are due to reverberation. 

For an irradiated disc, in general, we obtain a temperature profile shallower than the standard one 
$T \propto r^{-3/4}$ \citep{Sha73}. The reverberation hypothesis assumes that the optical/UV disc 
regions are irradiated by EUV/X-ray photons from the vicinity of the disc axis. If the high-energy 
source is placed on the axis and at a height $H_X$ above the thin disc (disc thickness $<< H_X$), 
the non-standard temperature profile is
\begin{equation}
T(r) = \left[\frac{3GM\dot{M}}{8\pi\sigma r^3} + \frac{(1-A)L_X H_X}{4\pi\sigma 
(H_x^2 + r^2)^{3/2}}\right]^{1/4} ,
\end{equation}
where $G$ is the gravitation constant, $\sigma$ is the Stefan constant, $\dot{M}$ is the mass 
accretion rate, $A$ is the disc albedo, i.e., the ratio of reflection to incident high-energy 
radiation, and $L_X$ is the luminosity of the irradiating source \citep[e.g.,][and references 
therein]{Cac07}. Moreover, considering $r >> H_X$ (and thus, a standard temperature profile), the 
typical radius of the intensity distribution at a given restframe wavelength $\lambda$ should be 
greater than the standard value \citep[for standard structure, see, e.g.,][]{Sha02}. Eq. (2) leads 
to 
\begin{equation}
R = \left[\frac{3GM\dot{M}}{8\pi\sigma} + \frac{(1-A)L_X H_X}{4\pi\sigma}\right]^{1/3} 
\left[\frac{k \lambda}{hc}\right]^{4/3} ,
\end{equation}
where $k$ is the Boltzmann constant, $h$ is the Planck constant, and $c$ is the speed of light. 
This non-standard typical radius is produced by both the heating due to irradiation and the viscous 
heating in the disc. Finally, we point out that shallow temperature profiles (from reverberation) 
could be consistent with microlensing data of some lensed quasars \citep[e.g.,][]{Poi08}. Moreover, 
the non-standard sizes of several lensed and microlensed quasars \citep{Mor07,Poo07} might be 
related to relatively high irradiation-to-viscosity ratios $IVR = 2(1-A)L_X H_X/3GM\dot{M}$.

\begin{acknowledgements}
We thank an anonymous referee for several comments that improved the presentation of our results. 
Liverpool Quasar Lens Monitoring is a long-term project to follow the optical variability of lensed 
quasars with the Liverpool robotic telescope. This paper is partially based on the results of the 
first phase of this project (LQLM I; see the Web site http://grupos.unican.es/glendama/index.htm). 
We acknowledge the continuing support of the Liverpool telescope team. We also thank T. Kundi\'c and
other members of the APO collaboration for providing light curves to us. This research has been 
supported by the Spanish Department of Education and Science grant AYA2007-67342-C03-02 and 
University of Cantabria funds. RGM holds a grant of the ESP2006-13608-C02-01 project financed by the 
Spanish Department of Science and Innovation.
\end{acknowledgements}

\end{document}